**Enhanced Raman scattering by fast GaN phonon-polaritons**

Mayssoune Mina, Toni Alhaddad, and Olivier Pagès[a)]

Université de Lorraine, LCP-A2MC, UR 201019679B, F-57000 Metz, France

[a)] Author to whom correspondence should be addressed : olivier.pages@univ-lorraine.fr

## Abstract

Phonon-polaritons propagating in crystal volume offer the possibility of transferring information throughout matter (via phonons) at high (photon-like) velocity and tunable frequency/wavelength in the far-infrared. However, from the phonon-polariton Raman cross section, the phonon and photon advantages seem mutually exclusive. Either the phonon-polaritons are fast (photon-like) but hardly supported by the lattice (not phonon-like), or they are well supported by the lattice (testified by a high Raman efficiency) but slow (not photon-like). An optimal phonon-photon coupling is currently searched for in hexagonal-GaN by near-forward Raman scattering across parallel crystal faces with in-plane singular $\vec{c}$-axis. Two accessible phonons, *i.e.*, the ordinary $A_1$ and $E_1$ ones, plus two refractive indices, *i.e.*, the ordinary and extraordinary ones, generate various phonon-polariton candidates. Notably, in perfect forward Raman scattering and in crossed polarizations of the incident ($\parallel \vec{c}$) and scattered lights, which, altogether, maximizes the transferred wavevector $\vec{q}$ to the crystal at minimum scattering angle, a fast phonon-polariton, stemming from the bottleneck of the $E_1$ dispersion on entry to the deep photon-like regime, is activated. It is well supported by the lattice since its Raman signal is strong and sharp, enhanced by multi-reflection of the laser beam between crystal faces at near-normal incidence. This fast Raman-enhanced phonon-polariton is interesting for infrared-photonics in that it cumulates the advantages of a photon (speed) and of a phonon (Raman intensity). Besides, it commutates from a phonon-polariton to a phonon by deviating from normal incidence or by permuting the incident and scattered polarizations, with potential applications as a vibrational/optical switch.

In polar semiconductor compounds with cubic (zincblende) or hexagonal (wurtzite) structure, like GaP and GaN, respectively, the optical lattice vibrations (phonons), resulting from opposite motion of the intercalated cation (Ga) and anion (P, N) sublattices, hence reflecting an effective bond stretching, are likely to carry an electric field due to the polarity of the cation-anion bonding.[1,2] The transverse optical (TO) vibration perpendicular to the direction of propagation (the latter being fixed by the phonon wavevector $\vec{q}$) is interesting for photonics because its electric field is transverse, *i.e.*, photon-like (for any $\vec{q}$-direction in a zincblende crystal and for $\vec{q}$ along or perpendicular to the singular $\vec{c}$-axis in a uniaxial wurtzite crystal). The resulting TO excitation with mixed phonon/photon character, called a phonon-polariton, offers a possibility of transferring information throughout matter (via phonons) at a high (photon-like) velocity in the far-infrared range ($\omega$) of lattice vibrations.[3] The bulk photon-polaritons propagating in the crystal volume attract special attention because they exhibit a dramatic $\omega$ vs. $q$ dispersion.[1-4] So, they can be tuned in frequency ($\omega$) or wavelength ($\lambda = 2\pi q^{-1}$) in view of targeted applications. In contrast, the surface phonon-polaritons, while less prone to energy dissipation by the lattice defects and impurities, are quasi dispersionless.[2,5]

However, there are limitations to the propagation of bulk phonon-polaritons. Out of the possible $\lambda$ values for a phonon spanned by the first Brillouin zone, from quasi-infinite ones near the Brillouin zone center ($q$=0) up to minimal ones of twice the lattice constant ($a$) near the Brillouin zone boundary, only those falling near the $\omega$ vs. $(q, \lambda)$ dispersion of a photon are eligible. A natural reference is the photon dispersion ($\omega = c \times q$) governed by the speed of light in vacuum ($c$). In matter, this is scaled down by the refractive index of the crystal ($n$) taking different values below ($n_s$, $s$ standing for "static", $\omega \ll \omega_{TO}$) and above ($n_\infty < n_s$, $\omega \gg \omega_{TO}$) the TO resonance. Both $(n_s, n_\infty)$-photon dispersions are quasi vertical, so that phonon-polaritons can only exist near the Brillouin zone center, in fact within the first per ten thousand ($10^{-4}$) of the Brillouin zone size ($\pi/a$).[2]

Experimentally, such small $q$ values are achievable by inelastic visible-light (Raman) scattering, that naturally operates near the Brillouin zone center by force of the quasi vertical dispersion of the visible laser excitation.[2] However, a near-perfect forward scattering geometry (schematically operating "in transmission"), minimizing the transferred wavevector $\vec{q}$ to the crystal, is required. Otherwise, the addressed $q$ cannot support the phonon-polariton coupling and the probed TO mode is deprived of electric field, reducing to a purely-mechanical TO phonon (PM-TO). Near-forward Raman studies of volume phonon-polaritons in pristine semiconductors were pioneered on cubic-GaP in the sixties.[6] Their latest development in the present century is concerned with hexagonal-GaN,[7-9] including pioneering *ab initio* calculations,[10] preceded by extended Raman[11-14] and *ab initio*[15] studies of the native PM-TO modes of the phonon-polaritons.

Even if an experimental access to volume phonon-polaritons is granted, their use in photonics is questionable. This is because the photon and phonon advantages seem mutually exclusive. Either the phonon-polariton is fast (photon-like) but hardly supported by the lattice (not phonon-like), or it is well anchored to the lattice (phonon-like) but slow (not photon-like). The amount of phonon character (and by complementarity of photon character) *per* phonon-polariton depending on $q$ (apparent, *e.g.*, in Fig. 2b of Ref. 2) can be estimated by looking at the phonon-polariton Raman cross section ($PP - RCS$), considering that only matter/phonon-like excitations, and not photon-like ones, scatter light efficiently. The GaN $PP - RCS$ shown in Fig. 1, presently calculated in a linear dielectric approach,[16] is illustrative with this respect. While large in the asymptotic *phonon* regimes at large and small $q$, the $PP - RCS$ turns small in the asymptotic *photon* regimes at large and small $\omega$. The best photon/phonon compromise is achieved in the strong phonon-polariton coupling regime manifested by the repulsion (anticrossing) of two coupled phonon-photons $PP^-$ and $PP^+$ modes (in order of frequency) at the crossing of the horizontal phonon asymptotes[17] and quasi vertical $(n_s, n_\infty)$-photon asymptotes.

$PP^+$ is hardly achieved by Raman scattering – but with exceptions.[18] Hence, $PP^+$ is not discussed hereafter. By contrast, $PP^-$ is accessible and there are positive signs, discussed below, of a real potential for $PP^-$ being used in infrared-photonics.

Recently, minor Mg-alloying of ZnTe proved successful in stabilizing a fast (photon-like) $PP^-$ originating from the highly dispersive photon-like bottleneck of the $PP^-$ dispersion and yet strongly bound to the lattice (phonon-like), testified by a high Raman intensity.[19] Basically, alloying introduces an intermediary $PP^{int}$ mode between $PP^-$ and $PP^+$ that couples with $PP^-$ via their common transverse electric field. Due to such coupling, $PP^-$ benefits from the strong phonon character manifested by $PP^{int}$ across most of its characteristic $S$-like dispersion.[20]

Such benefit due to alloying is absent in pristine crystals. Hence, the $PP^-$ Raman modes of cubic-GaP[6] and cubic-ZnS[21] gradually collapse on descending the $PP^-$ dispersion towards the photon regime. This is also true for $PP^-$ of hexagonal-GaN in most Raman scattering geometries explored by Torii *et al.*[7] and by Irmer *et al.*[8,9] addressing either the ordinary $A_1$ or $E_1$ phonon-polaritons polarized along and perpendicular to $\vec{c}$, respectively, or the extraordinary one with $A_1 - E_1$ mixed character. However, there is one intriguing example of a progressive enhancement of a sharp $E_1$-like $PP^-$ Raman mode ($PP_{E1}^-$) on penetrating the photon regime in Ref.[8] (Fig. 9 therein). This attracted no attention so far. Yet, a sharp and strong PP Raman signal in the photon-like regime means a long lifetime $PP_{E1}^-$ (sharpness) cumulating the advantages of a phonon (Raman intensity) and of a photon (speed).

In this work we investigate whether such photon/phonon-cumul for $PP_{E1}^-$ of hexagonal-GaN, so promising for infrared-photonics, is robust or accidental, with the ambition, in the robust case, to elucidate the mechanism behind the cumul.

In our near-forward Raman setup, the wavevector $\vec{q}$ transferred to the crystal, dictated by the impulse conservation $\vec{q} = \vec{k}_i - \vec{k}_s$, is monitored “from behind”, *i.e.*, by varying the laser incidence ($\vec{k}_i$) at the rear of the crystal, with fixed detection of the scattered light ($\vec{k}_s$) perpendicularly to the front crystal face. This deviates from earlier used setups in Refs.[7-9] in which $\vec{q}$ was adjusted the other way around, *i.e.*, “from the front” at normal laser incidence while positioning annular[7] or rectangular[8-9] apertures slightly off the laser axis in front of the detection lens. Using such apertures, the scattered light was collected at finite scattering angles ($\theta$) between $\vec{k}_i$ (fixed) and $\vec{k}_s$ (determined from the apertures) inside the crystal. In this way, the extraordinary phonon-polariton was successfully probed on top of the ordinary $A_1$ and $E_1$ ones.[7-9] In this work, we are mostly interested in $PP_{E1}^-$, as justified above. The $PP_{E1}^-$ Raman spectra are discussed by referring to the generic $(q, \theta)$-dependent $PP - RCS$ presently calculated for GaN within the linear dielectric formalism of Hon and Faust.[16,22]

Our GaN sample consists of a free-standing (0001)-oriented 5×5×0.4 mm$^3$ single crystal with out-of-plane $\vec{c}$-axis.[23] $PP_{E1}^-$ is accessed by forward Raman scattering across the edge crystal faces maintaining the in-plane $\vec{c}$-axis perpendicular to the incidence/detection ($\vec{k}_i, \vec{k}_s$)-plane (inset of Fig. 2). Besides $E_1$ and $A_1$, $E_2^H$ (polarized ⊥ $\vec{c}$) is also activated. In contrast with $E_1$ and $A_1$, $E_2^H$ is non-polar in character and thus cannot support the phonon-polariton coupling. Hence, $E_2^H$ emerges at the same frequency whether probed in forward or backward scattering. The same applies to the $E_1$ longitudinal optical ($LO_{E1}$) mode, also apparent. Indeed, its electric field being longitudinal, $LO_{E1}$ is immune to photonic effects.

The $\vec{k}_i$ and $\vec{k}_s$ directions are optimized by using long-focal (16 and 3 cm) lens to focus/collect the incident/scattered light at the rear/front of the crystal. In perfect forward Raman scattering ($\theta$=0°), $A_1$ and $E_1$ are separately addressed in crossed and parallel polarizations of the incident laser ($\vec{e}_i \parallel \vec{c}$) and of the scattered light ($\vec{e}_s$), *i.e.*, in the $x(z,y)x$ and $x(z,z)x$ scattering geometries using the compact Porto's notation $\vec{k}_i(\vec{e}_i, \vec{e}_s)\vec{k}_s$,[24] respectively, with $(x, y, z \parallel c)$ denoting the crystal axes. Special attention is awarded to $PP_{E1}^-$. This is studied, firstly, in its $\theta$-dependence in crossed ($\vec{c} \parallel \vec{e}_i \perp \vec{e}_s$) polarizations with various (blue: 488.0 nm, red: 632.8 nm, near-infrared: 785.0 nm) laser excitations (Fig. 2),[25] and, secondly, at minimum $\theta$ with the blue laser line in crossed polarizations by varying the azimuth angle $\alpha$ between $\vec{e}_i$ and $\vec{c}$, step-increased by 20° until completing a full revolution at a given crystal spot (Fig. 3). All Raman spectra are acquired in the Stokes geometry. The scattered light is less energetic than the incident laser ($\omega_s < \omega_i$) in the energy conservation ($\omega = \omega_i - \omega_s$), featuring an effective transfer of energy ($\omega$) to the crystal.

Forward $E_1$ Raman spectra obtained at decreasing incidences of the blue laser in crossed polarizations ($\vec{c} \parallel \vec{e}_i \perp \vec{e}_s$) are shown in Fig. 2. $PP_{E1}^-$ progressively strengthens until a sharp and strong Raman signal at 485 cm$^{-1}$ emerges at minimal scattering angle ($\theta_{min}$). This echoes earlier results of Irmer *et al.*[8] (see above). The $PP_{E1}^-$ Raman enhancement at $\theta \rightarrow 0$ is thus robust and not accidental.

Its explanation requires in the first place an accurate $\theta_{min}$ estimate. This is achieved by carefully adjusting $\theta$ until the corresponding $\omega(q,\theta)$ Raman scan line for the blue excitation crosses the $PP_{E1}$-$RCS(q,\omega)$ (Fig. 1, yellow curves) exactly at the experimental $PP_{E1}^{-}$ frequency (~485 $cm^{-1}$). In this way, we find $\theta_{min}$~0.8°, suggesting that the $PP_{E1}^{-}$ enhancement results from multi-reflection of the laser beam between parallel crystal faces near normal incidence ($\theta$~0°). In fact, the $PP_{E1}^{-}$ enhancement is consistently repeated at $\theta$~0° whichever laser line is used (Fig. 1), *i.e.*, blue (a), red (b) or near-infrared (c). Corresponding series of $\theta$-dependent Raman spectra are shown in Fig. S1 (the prefix $S$ stands for supplementary information) completing Fig. 1.

Technically, the Raman scan line $\omega(q)$ for a given $\theta$ is derived from the impulse conservation $\vec{q} = \vec{k}_i - \vec{k}_s$ by expressing $k_{i,s}$ in the form $c^{-1} \times \omega_{i,s} \times n(\omega_{i,s})$, with $n(\omega_{i,s})$ representing the refractive index of GaN at the laser frequency ($\omega_i$) and at lower ($\omega_s$) frequencies (Stokes scattering). $q$=$\sqrt{\vec{q}^2}$ is univocally related to $\theta$ entering in the scalar product between $\vec{k}_i$ and $\vec{k}_s$. In the used ($\vec{c} \parallel \vec{e}_i \perp \vec{e}_s$) polarization setup, the refractive indices for the incident laser and scattered light are the extraordinary $n_e(\omega_i)$ and ordinary $n_o(\omega_s)$ ones, respectively, taken from Ref. 26. As for the $PP_{E1}$-$RCS(q,\omega)$, the GaN input parameters are the $E_1$-like Faust-Henry coefficient ($C_{F-H}$ =–2.31), the PM-TO frequency ($\omega_{TO,E1}$ =558.8 $cm^{-1}$) and the static ($\varepsilon_s$ =9.12) and high-frequency ($\varepsilon_\infty$ =5.19) relative dielectric constants (referring to $\vec{q} \perp \vec{c}$, the studied case in this work), taken from Refs. 8 and 9.

By reducing the laser energy at $\theta$~0°, the enhanced $PP_{E1}^{-}$ shifts to lower frequency and broadens.

The shift relates to the achieved $q_{min} = c^{-1}|n_e(\omega_i) \times \omega_i - n_o(\omega_s) \times \omega_s|$ value at $\theta$~0°. In the used polarization setup ($\vec{c} \parallel \vec{e}_i \perp \vec{e}_s$), $q_{min}$ remains finite because the positive difference in energy $\omega_i - \omega_s$ (Stokes scattering) is accentuated by the positive difference in refractive index $n_e - n_o$ (GaN is a positive uniaxial crystal).[26] The dispersion of the GaN refractive indices, however, reduces with energy,[26] with concomitant impact on $n_e - n_o$, being also reduced. So, the achieved $q_{min}$ at $\theta$~0° gets down by using less energetic laser excitations, penetrating deeper downward the $PP_{E1}^{-}$ dispersion towards low frequency.

The broadening relates to the finite angles of the incidence ($\vec{k}_i$) and detection ($\vec{k}_s$) cones. Due to such experimental limitations, a finite $\theta$-domain is probed instead of a unique $\theta$. Near the normal incidence, $\theta$~0° enters the probed $\theta$-domain as the smallest achievable $\theta$. The related Raman signal is enhanced by multi-reflection, marking the $PP_{E1}^{-}$-maximum. The enhancement is reduced on departing from $\theta$~0°. The enhanced-$PP_{E1}^{-}$ Raman signal is thus asymmetrical towards high-frequency. The $\theta$-domain, defined by the extreme $\theta$ corresponding to maximum and minimum $PP_{E1}^{-}$ intensities (horizontal dotted lines, Fig. 1), hardly changes with the laser line, *i.e.*, $\Delta\theta$~1.5°±0.5°. Yet, the enhanced-$PP_{E1}^{-}$ broadens by reducing the laser energy, *i.e.*, by decreasing $q_{min}$. This is because the $PP_{E1}^{-}$ dispersion becomes so steep on penetrating downward the deep photon-like regime.

In the used ($\vec{c} \parallel \vec{e}_i \perp \vec{e}_s$) polarization setup, only $TO_{E1}$ polarized along $y$ is allowed at normal incidence/detection along $x$,[27] in the form of either $PM - TO_{E1}$ in backward scattering ($\theta$~180°) or $PP_{E1}^{-}$ in forward scattering ($\theta$~0°). $TO_{A1}$ and $LO_{E1}$, polarized along $z \parallel c$ and $x$, respectively, are forbidden. Hence, there is no doubt that the enhanced Raman signal at $\theta$~0° (Figs. 2 and 3) is $PP_{E1}^{-}$. Note that $PP_{E1}^{-}$ generated in forward scattering is accompanied by various PM-TO modes generated in backscattering after reflection of the laser beam at the front crystal face. Both the $PP_{E1}^{-}$ and PM-TO Raman signals are enhanced near normal incidence. As soon as deviating from the strict incidence/detection along $x$, the Raman selection rules are relaxed and $PP_{A1}^{-}$ becomes visible besides $PP_{E1}^{-}$ at large scattering angle ($\theta \neq 0°$, Fig. 2).

The enhanced-$PP_{E1}^{-}$ at $\theta$~0° with the blue laser line (corresponding to the sharpest signal) is further studied in its dependence on the azimuth $\alpha = \widehat{(\vec{e}_i, \vec{c})}$ of the crossed ($\vec{e}_i \perp \vec{e}_s$) polarizations rotated in pair as a solid body until completing a full revolution at the same sample spot (Fig. 3). The theoretical pattern featuring the $\alpha$-dependence of the $TO_{E1}$ Raman intensity, dictated by $|\vec{e}_i R_y \vec{e}_s|^2$ with $R_y = \begin{pmatrix} \cdot & \cdot & \cdot \\ \cdot & \cdot & c \\ \cdot & c & \cdot \end{pmatrix}$ representing the Raman tensor for a mode polarized along $y$,[7-9] contains four symmetrical branches axed $y$ and $z$ (upper inset of Fig. 3, solid curve). Only the $z$-branches are apparent for $PP_{E1}^{-}$ (symbols); the $y$-branches are "missing". In contrast the native $PM - TO_{E1}$ behind $PP_{E1}^{-}$, generated in backscattering after reflection of the laser beam at the front crystal face exhibits the four nominal maxima along $y$ and $z$ (marked by stars).

To our view, the "ghost" $y$-branches for $PP_{E1}^{-}$ are not really "missing", but merely "channeled" towards $LO_{E1}$. In fact, the $PP_{E1}^{-}$ and $LO_{E1}$ Raman intensities exhibit antagonist $\alpha$-variations in Fig. 3, a maximum for one corresponding to a minimum for the other. Such antagonism clarifies by comparing the $\vec{q}_{min}$ situations at $\theta \sim 0°$ at two representative $\alpha$, *i.e.*, $\alpha_1 = 0°$ ($PP_{E1}^{-}$ active, $LO_{E1}$ quasi inactive) and $\alpha_2 = 90°$ ($PP_{E1}^{-}$ inactive, $LO_{E1}$ active).

As already discussed, $q_{min}$ achieves maximum at $\alpha_1$, due to a favorable scaling of the $n_e$ and $n_o$ refractive indices attached to the incident laser and to the scattered light, respectively. The scaling is reversed at $\alpha_2$ due to permutation of the incident and scattered polarizations. In this case, the positive difference in energy $\omega_i - \omega_s$ (Stokes process) is compensated by the negative difference $n_o - n_e$ between refractive indices. Consequently, $q_{min} \to 0$ at $\alpha_2$, and, consequently, $k_i \sim k_s$. Strictly at $\theta$=0°, $\vec{q}_{min}$ is parallel to $x$. The Raman active mode polarized along $y$ vibrating perpendicularly to $\vec{q}_{min}$ (transverse character) and to $z \parallel c$ ($E_1$ character) is $PP_{E1}^{-}$. However, the perfect forward Raman scattering is never achieved experimentally, as already discussed (recall the finite $\Delta\theta$-domain). The minimum achievable $\theta$ value, while remaining small, *i.e.*, $\sim$1.5° in our case (by averaging over the mean $\theta$ values of the found $\Delta\theta$-domains for the blue, red and near-infrared excitations, Fig. 1) is actually finite. This, in fact, has dramatic impact on the $\vec{q}_{min}$-orientation at $\alpha_2$. Being small, $\vec{q}_{min}$ is dominantly aligned with $y$ (Fig. 3, upper inset). In this case the Raman-active mode polarized along $y$ vibrates along $\vec{q}_{min}$ (longitudinal character) perpendicularly to $z \parallel c$ ($E_1$ character), identifying with $LO_{E1}$. Hence, by varying $\alpha$, the Raman-active $y$-polarized $E_1$ mode probed in forward Raman scattering periodically oscillates between $PP_{E1}$ ($\alpha_1$-like) and $LO_{E1}$ ($\alpha_2$-like).

Retrospectively, a similar interplay between $PP_{E1}^{-}$ (enhanced at $\theta \searrow$) and $LO_{E1}$ (enhanced at $\theta \nearrow$) is apparent at fixed ($\vec{c} \parallel \vec{e}_i \perp \vec{e}_s$) polarizations by varying $\theta$ (Fig. 2). In this polarization setup, $k_i > k_s$, as already discussed. Near $\theta_{min}$, $\vec{q} = \vec{k}_i - \vec{k}_s$ is nearly parallel to $x$ and the active Raman mode determined by $R_y$ is $PP_{E1}^{-}$ (lower inset). By increasing $\theta$, $\vec{q}$ develops a $y$-component that, from a certain $\theta$, *i.e.*, $\sim$4° here (Fig. 1), becomes prominent. At this limit, the Raman-active $y$-mode $E_1$ is mainly polarized along $\vec{q}$, hence longitudinal, identifying with $LO_{E1}$ (upper inset). In a nutshell, $E_1$ manifests as the transverse-$PP_{E1}^{-}$ at $\theta \to 0$ and as the longitudinal-$LO_{E1}$ at large $\theta$, leaving the impression in Fig. 2 that $PP_{E1}^{-}$ is progressively relayed by $LO_{E1}$ by departing from $\theta \sim 0°$.

Last, one may well wonder why only $PP_{E1}^{-}$, and not $PP_{A1}^{-}$, is enhanced at $\theta \to 0$. $PP_{A1}^{-}$ is Raman-active in parallel ($\vec{c} \parallel \vec{e}_i \parallel \vec{e}_s$) and crossed ($\vec{e}_i \perp \vec{e}_s$)-polarizations at $\alpha = 0°$ and 45°, respectively. In both cases, the same refractive index is used for the incident laser and for the scattered light. The corresponding $\theta = 0°$ Raman scan line with the most energetic (blue) laser line probes the $PP_{A1}^{-}$ dispersion extremely deep into the photon-like regime, *i.e.*, below 100 cm$^{-1}$, to compare with 485 cm$^{-1}$ for $PP_{E1}^{-}$ in similar conditions (as apparent at $\alpha = 0°$ in Fig. S4). At this limit the $PP_{A1}^{-} - RCS$ vanishes to nearly zero, being hardly enhanced by multi-reflection of the laser beam between crystal faces at $\theta \sim 0°$. The situation gets even worse by using the less energetic red and near-infrared excitations that penetrate deeper downward the $PP_{A1}^{-}$ dispersion. Hence, no chance is given in the $A_1$ symmetry to benefit from the PP-Raman enhancement at $\theta \sim 0°$.

Summarizing, $PP_{E1}^{-}$ of GaN detected in near-perfect forward Raman scattering in $\vec{c} \parallel \vec{e}_i \perp \vec{e}_s$ crossed-polarizations is interesting in infrared-photonics in that it cumulates the advantages of a photon (speed) and of a phonon (lattice anchoring). First, it stems from the bottleneck of the $PP_{E1}^{-}$ dispersion on entry of the photon-like regime, hence fast. Second, it is well supported by the lattice (phonon-like), testified by a large Raman efficiency, enhanced by multi-reflection of the laser beam between crystal faces. Besides, $PP_{E1}^{-}$ transforms from a (transverse) phonon-polariton to a (longitudinal) phonon, *i.e.*, $LO_{E1}$, by permuting polarizations or by slightly deviating from normal incidence, offering a versatile and sensitive vibrational/optical switch.

## SUPPLEMENTARY MATERIAL

Supplementary material serves several purposes. First, the $\theta$-dependent series of cross-polarized $PP_{E1}$ near-forward Raman spectra behind the reported data in Fig. 1, corresponding to various laser excitations, are displayed in Fig. S1. Second, the exact scattering geometries behind those sketched out in the central/lower insets of Figs. 2 and 3 are reported in Figs. S2 and S3, respectively. Last, the $q$-dependent $PP_{A1}$ Raman cross section of hexagonal-GaN is introduced in Fig. S4, for comparison with $PP_{E1}$.

## ACKNOWLEDGEMENTS

This work was supported by the *French PIA project Lorraine Université d'Excellence, part of the France 2030 Program, reference ANR-15-IDEX-04-LUE*, within the ViSA – IRP «**Vi**brations of **S**emiconductor **A**lloys – **I**nternational **R**esearch **P**artnership» wall-less associated international laboratory (2024 – 2028), co-funded by the Excellence Initiative – Research University program at Nicolaus Copernicus University in Toruń. We acknowledge technical assistance of Pascal Franchetti in performing the Raman measurements.

## AUTHOR DECLARATIONS

### Conflict of Interest

The authors have no conflicts to disclose.

## AUTHOR CONTRIBUTIONS

**Mayssoune Mina:** Formal analysis (equal); Investigation (equal); Software (equal); Validation (equal); Visualization (lead). **Toni Alhaddad:** Formal analysis (equal); Investigation (equal); Software (equal); Validation (equal). **Olivier Pagès:** Conceptualization (lead); Funding acquisition (lead); Methodology (lead); Project administration (lead); Supervision (lead); Visualization (equal); Writing – original draft preparation (lead).

## DATA AVAILABILITY

The data that support the findings of this study are available from the corresponding author upon reasonable request.

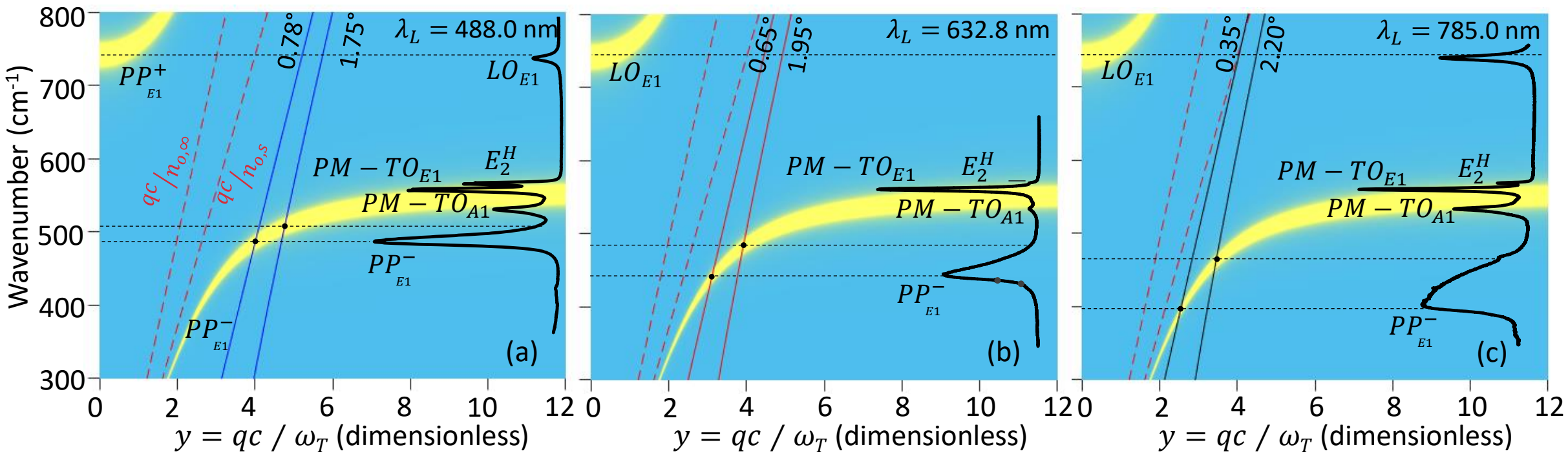

**FIG. 1. $E1$ phonon-polariton of hexagonal-GaN in its $q$-dependence.** Dispersion (plain yellow curves) and Raman intensity (thickness of plain yellow curves) generated from the $q$-dependent $PP_{E1} - RCS$. The $n_O$-related photon asymptotes are indicated (dashed/red lines). The $PP_{E1}^{-}$ Raman signals (left-oriented solid/black curves) recorded at near-normal incidence ($\theta \sim 0°$, as indicated) with the (a) 488.0 nm, (b) 632.8 nm and (c) 785.0 nm laser lines cover finite $\theta$-domains. The Raman scan lines (solid/colored curves) referring to the extreme $\theta$ values (as specified) corresponding to maximum and minimum $PP_{E1}^{-}$ Raman intensity (horizontal dotted lines) are indicated. $\omega_T$ involved in the dimensionless $y$ parameter (abscissa axis) is the $PM - TO_{E1}$ frequency.

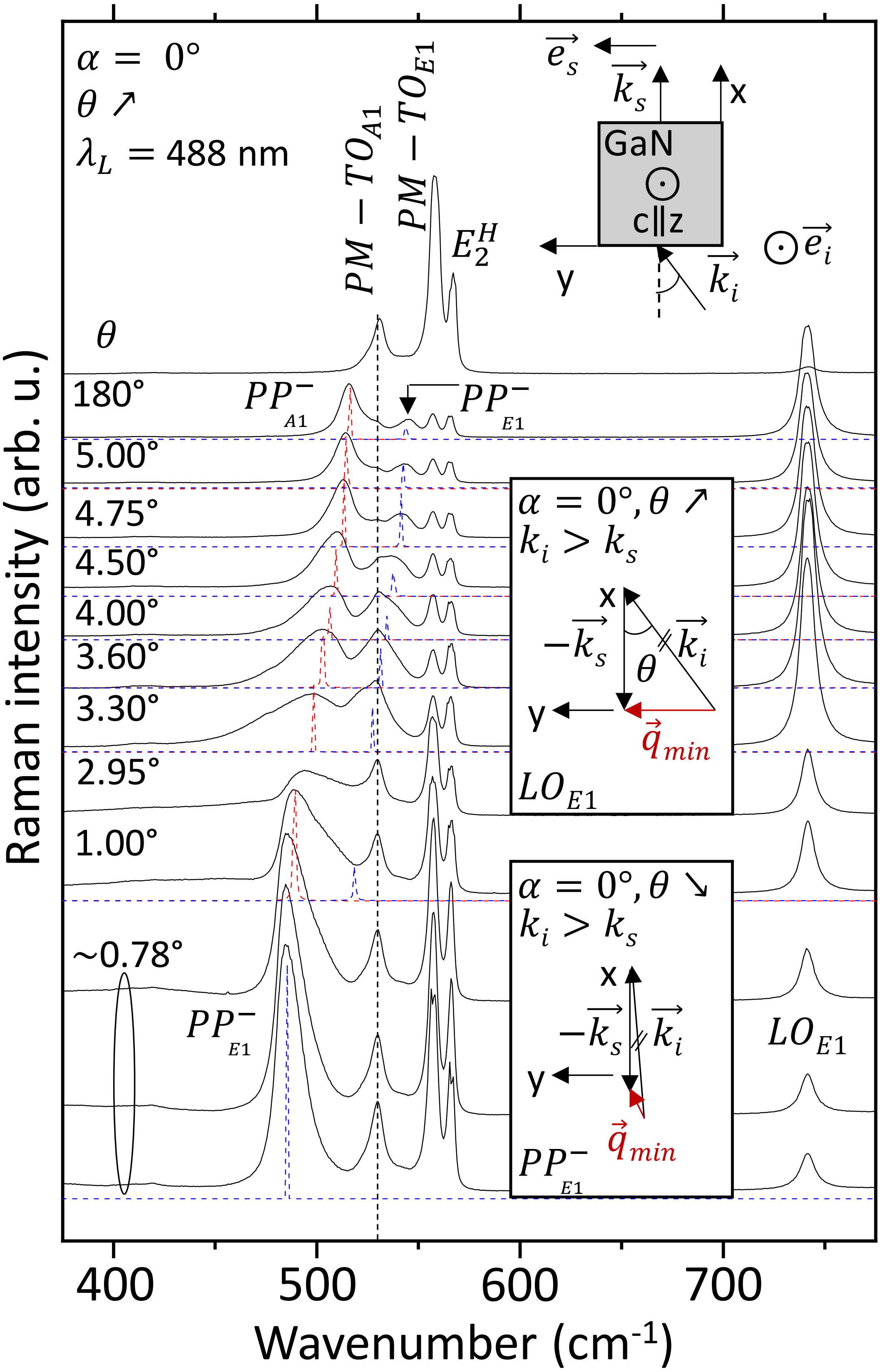

**FIG. 2. $\theta$-dependent cross-polarized $E_1$ near-forward Raman spectra of hexagonal-GaN.** In crossed-polarizations (upper inset, $\alpha = 0°$ – see Fig. 3), large and small scattering angles $\theta$ generate different scattering geometries activating $TO_{E1}$ in the form of either $LO_{E1}$ (central inset) or $PP_{E1}^{-}$ (lower inset). The exact scattering geometries are shown in Fig. S2. Dashed curves mark maxima in the $PP_{A1} -$ (red) and $PP_{E1} - RCS$ (blue).

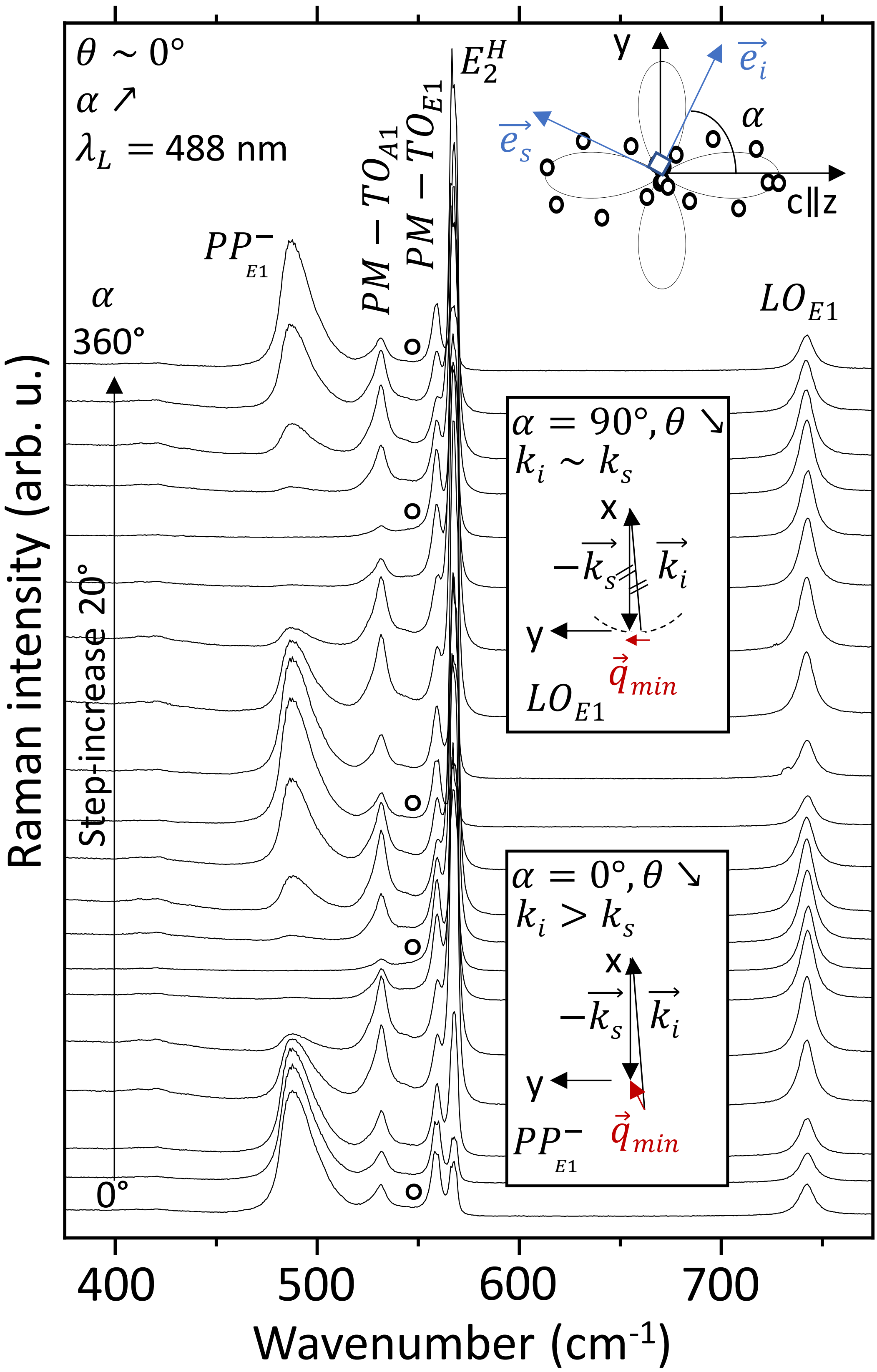

**FIG. 3. Cross-polarized near-forward Raman spectra of hexagonal-GaN at $\theta \sim 0°$ in their $\alpha$-dependence.** $PP_{E1}^{-}$ (symbols, upper inset) does not follow the nominal $TO_{E1}$ Raman selection rules (solid curve, upper inset), in contrast with its native $PM - TO_{E1}$, as emphasized by the circle. $TO_{E1}$ oscillates between its $PP_{E1}^{-}$ and $LO_{E1}$ forms due to a change in scattering geometry when $\alpha$ varies, as sketched out (lower and central insets). The exact scattering geometries are shown in Fig. S3.

**Supplementary material for "Enhanced Raman scattering by fast GaN phonon-polaritons"**

Mayssoune Mina, Toni Alhaddad, and Olivier Pagès[a)]

*LCP-A2MC, UR 4632, Université de Lorraine, 57000 Metz, France*

[a)] Author to whom correspondence should be addressed : olivier.pages@univ-lorraine.fr

Supplementary material serves several purposes. In **Sec. SI**, we report on the $\theta$-dependent series of cross-polarized $PP_{E1}$ near-forward Raman spectra behind the reported data in Fig. 1, corresponding to various laser excitations (Fig. S1). In **Sec. II**, we show the exact scattering geometries (Figs. S2 and S3) behind those sketched out in the central/lower insets of Figs. 2 and 3. Last, in Sec. III, we introduce the $q$-dependent $PP_{A1}$ Raman cross section of hexagonal-GaN (Fig. S4), for comparison with $PP_{E1}$.

## **I.** $\theta$-dependent near-forward $PP_{E1}^{-}$ Raman spectra of hexagonal-GaN ($\alpha = 0°$)

To complete Fig. 2 reporting on the cross-polarized ($\alpha = 0°$, as sketched out in the upper inset) near-forward Raman spectra of hexagonal-GaN in their $\theta$-dependence taken with the blue (488.0 nm) laser line, we provide in Fig. S1 similar data obtained with the (a) red (632.8 nm) and (b) near-infrared (785.0 nm) excitations. The exact scattering angle $\theta$ per spectrum and per excitation is estimated by adjusting the relevant $\theta$-dependent Raman scan line until this crosses the $PP_{E1}^{-}$ dispersion exactly at the observed Raman frequency, as apparent in Fig. 1.

## **II.** Near-forward GaN Raman scattering geometries depending on $\theta$ and $\alpha$

The exact scattering geometries behind those sketched out in Figs. 2 and 3 (in the central and lower insets) are represented in Figs. S2 ($\alpha$=0°, $\theta$ variable) and S3 ($\alpha$=0° and 90°, $\theta = \theta_{min}$) by involving the proper $\vec{k}_i$ and $\vec{k}_s$ magnitudes and orientations at the probed $PP_{E1}^{-}$ ($\omega_s$) frequencies and the proper scattering angles $\theta$ in various $\alpha$-dependent crossed-polarizations setups.

## **III.** $PP_{A1}$ Raman cross section of hexagonal-GaN in its $q$-dependency

The $q$-dependent $PP_{A1}$ Raman cross section of hexagonal-GaN calculated within the linear dielectric formalism set by Hon and Faust is shown in Fig. S4.[16] Selected $n_e$-related (involved in case of parallel polarizations along $\vec{c}$) Raman scan lines at minimal scattering angles ($\theta$=0°, 1°) penetrate deep downward the photon-like regime where the dependent $PP_{A1}^{-}$ Raman cross section vanishes to zero. This suppresses, in fact, any possibility of a strong enhancement of the $PP_{A1}^{-}$ Raman signal at $\theta$~0°, contrasting with experimental observations on $PP_{E1}^{-}$ in similar conditions (see main text).

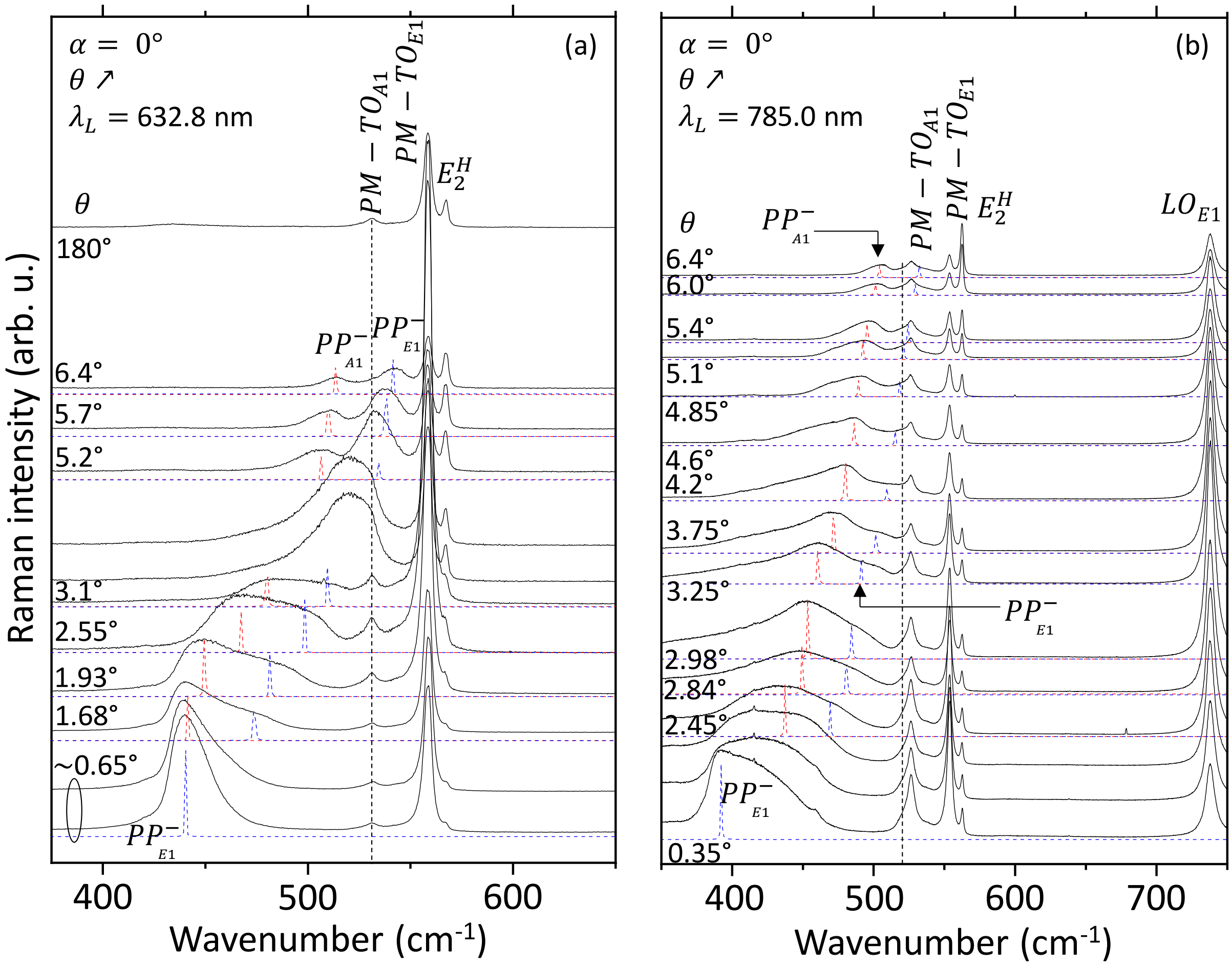

**FIG. S1. $\theta$-dependent cross-polarized $E_1$ near-forward Raman spectra of hexagonal-GaN taken with low-energy laser excitations.** Cross-polarized ($\alpha = 0°$, refer to the upper inset of Fig. 3) near-forward Raman spectra of GaN in the dominant $E_1$ symmetry taken at increasing scattering angle $\theta$, as specified, with the (a) red (632.8 nm) and (b) near-infrared (785.0 nm) laser excitations. Such data complete the similar series of spectra taken with the blue laser line, displayed in Fig. 2. Dashed curves mark maxima in the $PP_{A1} -$ (red) and $PP_{E1} - RCS$ (blue).

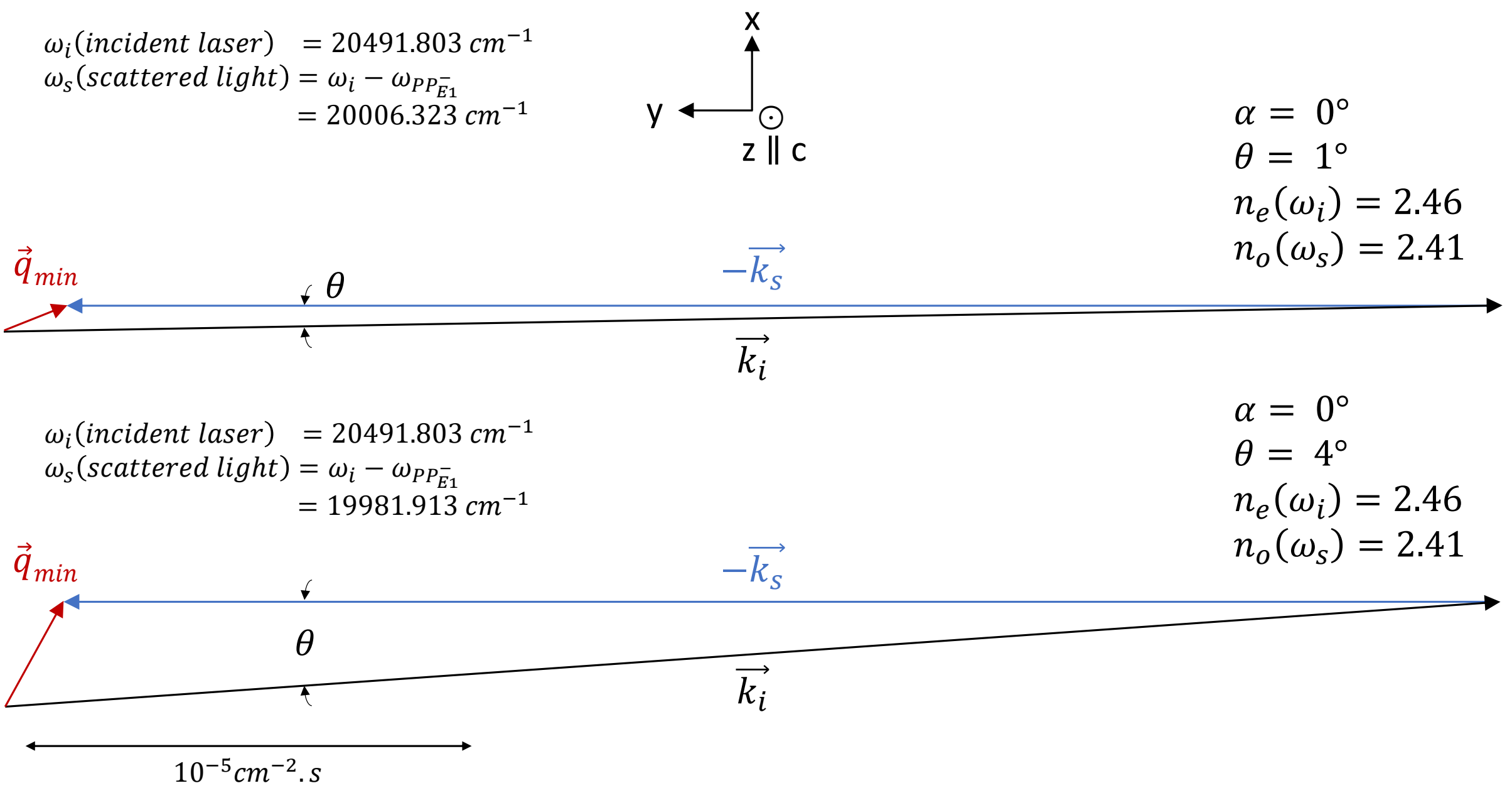

**FIG. S2. Near-forward $E_1$ scattering geometries in hexagonal-GaN ($\alpha$=0°, $\theta$ ↗).** Properly scaled ($\vec{k}_i$, $\vec{k}_s$) scattering geometries (in magnitude and orientation) involved in crossed-polarizations of the incident laser and of the scattered light ($\alpha$=0°, refer to the upper inset of Fig. 3) at increasing scattering angle $\theta$, as specified, corresponding to the reported near-forward Raman GaN spectra in Fig. 2. $\vec{x} \parallel \vec{k}_s$ is normal to the rear and front crystal faces. The used parameters to estimate $k_{i,s}$ are indicated.

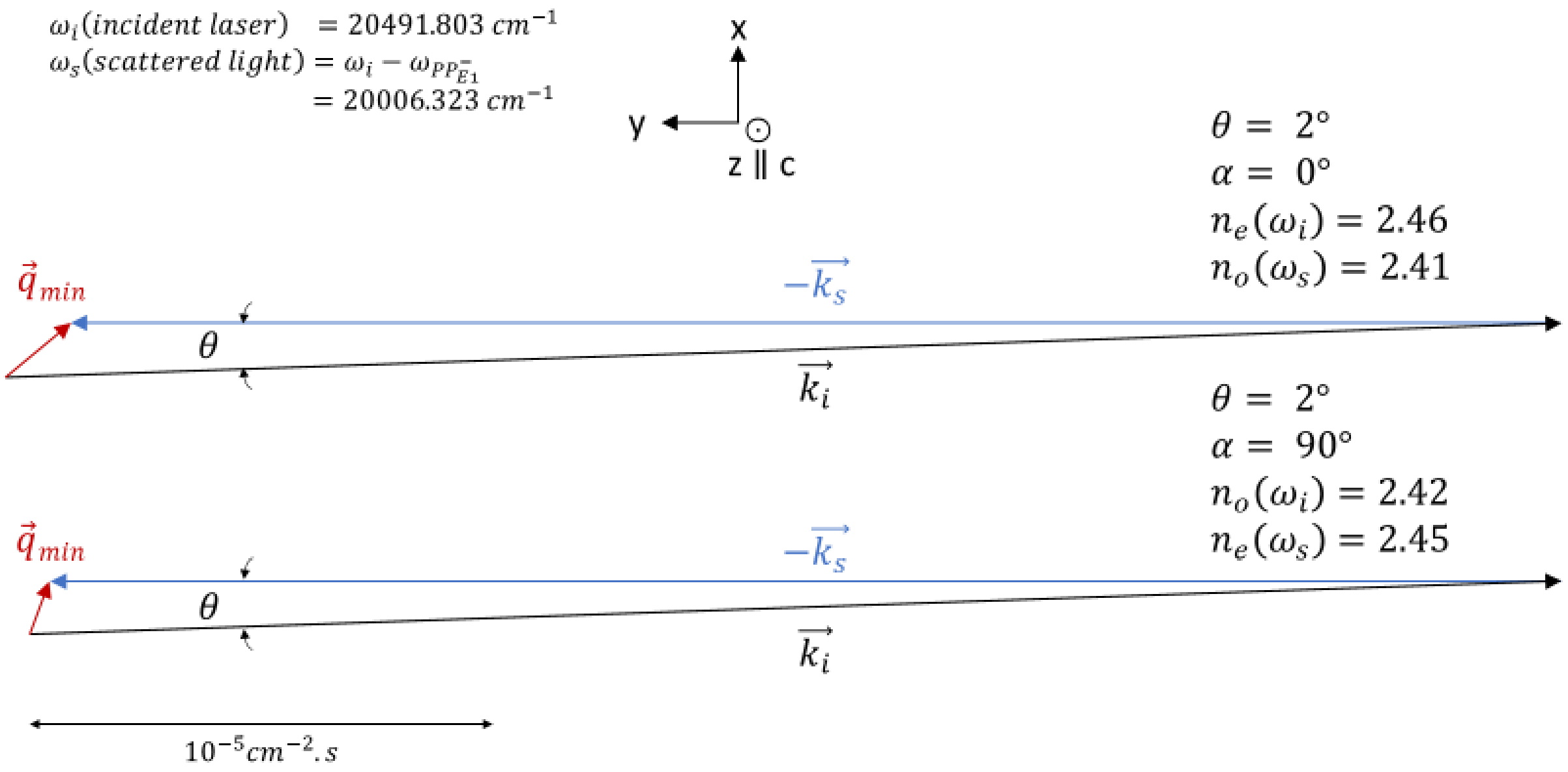

**FIG. S3. Near-forward $E_1$ scattering geometries in hexagonal-GaN ($\alpha$=0° and 90°, $\theta = \theta_{min}$).** Properly scaled ($\vec{k}_i$, $\vec{k}_s$) scattering geometries (in magnitude and orientation) involved at minimal scattering angle ($\theta = \theta_{min}$~2°) in crossed-polarizations of the incident laser and of the scattered light in their $\alpha$-dependence (refer to the upper inset of Fig. 3), as specified, corresponding to the reported near-forward Raman GaN spectra in Fig. 3. $\vec{x} \parallel \vec{k}_s$ is normal to the rear and front crystal faces. The used parameters to estimate $k_{i,s}$ are indicated.

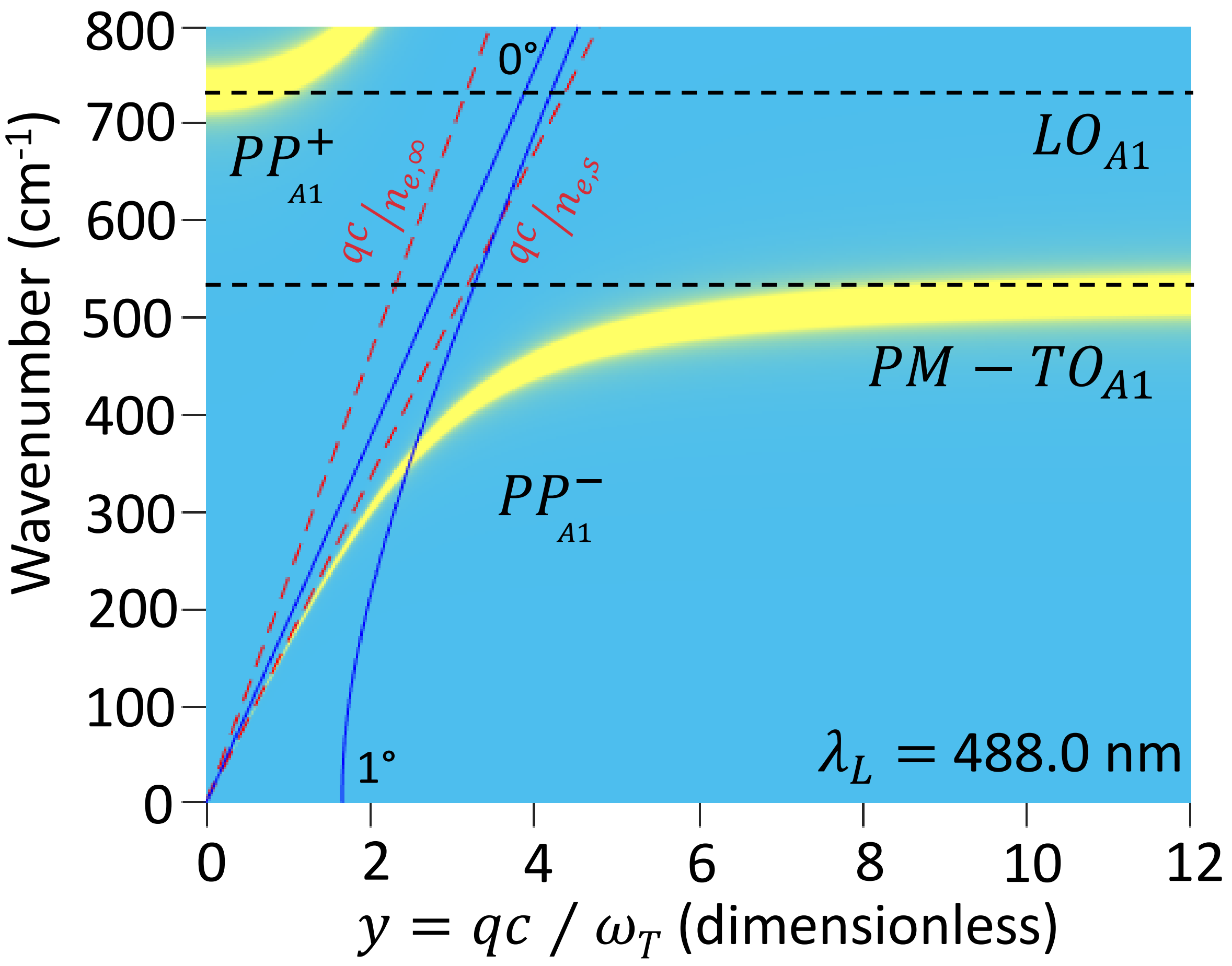

**FIG. S4. $PP_{A1}$ Raman cross section of hexagonal-GaN in its $q$-dependence.** Dispersion (plain yellow curves) and Raman intensity (thickness of plain yellow curves) generated from the $q$-dependent $PP_{A1}$ Raman cross section. The $n_e$-related photon asymptotes are indicated (dashed/red lines). The $n_e$-related (involved in case of parallel polarizations along $\vec{c}$) Raman scan lines (solid/blue curves) addressed with the 488.0 nm laser line at small/representative $\theta$ values (as specified) are indicated. $\omega_T$ involved in the dimensionless $y$ parameter (abscissa axis) is the $PM-TO_{A1}$ frequency, *i.e.*, 531.8 cm$^{-1}$.[8] The Faust-Henry coefficient for $A_1$ is $C_{F-H}$ =–3.81.[9]